# Giantically blue-shifted visible light in femtosecond mid-IR filament in fluorides

**A.E. Dormidonov,[1*] V.O. Kompanets,[2] S.V. Chekalin,[2] and V.P. Kandidov[1]**

[1]*Moscow Lomonosov State University, Physics Department and International Laser Center, 119991, Moscow, Russia*
[2]*Institute of Spectroscopy RAS, 142190, Moscow, Troitsk, Russia*
**dormidonov@gmail.com*

**Abstract:** A giant blue shift (more than 3000 nm) of an isolated visible band of supercontinuum was discovered and studied in the single filament regime of Mid-IR femtosecond laser pulse at powers slightly exceeding critical power for self-focusing in fluorides. At the pulse central wavelength increasing from 3000 nm to 3800 nm the spectral maximum of the visible band is shifted from 570 nm and 520 nm up to 400 nm and 330 nm for $BaF_2$ and $CaF_2$, respectively, its spectral width (FWHM) being reduced from 50 – 70 nm to 14 nm. It is shown that the formation of this narrow visible wing is a result of the interference of the supercontinuum components in the anomalous group velocity dispersion regime.

## 1. Introduction

Nonlinear transformation of a femtosecond pulse spectrum in a filament offers a new, unique tool for time-resolved spectroscopy. Formation of frequency-angular spectrum of filament supercontinuum (SC) was studied in detail for near-infrared pulses [1,2]. For Ti:Sapphire laser pulses with the central wavelength in the region of normal group-velocity dispersion (GVD) of fused silica and of many other condensed media the spectral broadening is monotonous in Stokes and anti-Stokes directions [3]. Three-octave-spanning SC generation in normal GVD regime at 2100 nm was demonstrated in ZnS [4]. For pulses with peak power much greater than the critical power for self-focusing SC spectrum is asymmetrically broadened to short-wave region in BaF$_2$ [5], fused silica, and LiF [6]. Filamentation of pulses in anomalous GVD regime leads to the formation of maximum in the visible spectral range [7–10]. Transformation of the SC spectra under filamentation of near-infrared femtosecond pulses at central wavelengths in the range 1300 – 2200 nm in fused silica was experimentally and numerically studied in [11–13]. It was found that in the IR pulse filament the anomalous GVD of fused silica leads to formation of an isolated anti-Stokes wing (ASW), which is located in the visible region of SC. With increasing centre wavelength from 1300 to 2300 nm, the spectrum of the ASW narrows, shifting to the blue. It was shown that SC short-wavelength cutoff and anti-Stokes shift are determined by multiphoton order of laser plasma generation process, a broad minimum in the SC spectrum, separating the anti-Stokes wing from the centre wavelength, is a result of destructive interference of the broadband SC. Observation of blue-shifted continuum peak in the visible region even when the filament is formed by near-infrared pulses has been experimentally confirmed in [14]. The authors identified the mechanisms responsible for the generation of this peak, clearly separated from the SC around the pump wavelength, as three-wave mixing. Direct connection of SC blue wing with near IR light bullets (LB) formation has been established in [15–17]. It was found that the energy of the SC visible part increases discretely by equal portions according to the number of LBs in the anomalous GVD filament. The LB is a stable self-organizing formation that doesn't depend on the wide range of the input pulse parameters. Recently LBs and SC spectrum transformation under variation of nonlinear interaction length of 1800 nm femtosecond pulse in filament have been investigated by focusing laser pulse at the front face of wedge-shaped fused silica sample [18] according to the scheme proposed in [19].

With a creation of femtosecond Mid-IR laser sources [20] studies of pulse filamentation at corresponding spectral range have been started. The possibility of LB formation by Mid-IR laser pulses propagating in humid air has been numerically investigated in [17,21,22]. Recently filamentation of ultrashort 3.9-μm pulses with energies above 20 mJ in the atmosphere has been demonstrated and the generation of powerful remarkably broad supercontinuum, stretching from the visible to the mid-infrared has been revealed [23]. A broadband multi-octave SC extending from 450 nm to 4500 nm was registered during the filamentation of pulses at the central wavelength of 3100 nm in YAG crystal [24]. The filamentation without fundamental spectrum broadening and SC generation was reported in CLEO:2013 [25], where UV-emission ascribed to F-centers was observed in YAG, CaF$_2$, Sapphire, and LiSGaF under the action of femtosecond pulses at 3900 nm.

It should be noted that fluorides are very interesting materials for filamentation investigation in Mid-IR, because their transparency band overlaps the UV and IR ranges that is important for SC formation and their GVD is anomalous that is necessary for LBs formation. Such an investigation can give a lot of important information on peculiarities of SC and LB formation in transparent media under filamentation of powerful femtosecond pulses in anomalous GVD conditions.

In this Letter we report experimental, numerical, and analytical studies of Mid-IR femtosecond laser pulse filamentation at central wavelengths near 3000 nm in BaF$_2$ and CaF$_2$. It has been revealed that plasma and LB formation in filament results in narrowband conical emission (CE) in the visible range. For the first time in Mid-IR spectral range a giant blue shift of an isolated spectral band of visible light was registered and studied with increasing of the initial pulse central wavelength. The formation of this narrow isolated blue shifted band is explained on the base of interference model.

## 2. Methods

The pulses at wavelengths of 2500 – 3800 nm, corresponded to the regions of anomalous dispersion in both CaF$_2$, and BaF$_2$ (zero GVD at 1546 nm and 1925 nm, correspondingly) were generated by a travelling-wave optical parametric amplifier of white-light continuum (TOPAS-C) with a noncollinear difference frequency generator (NDFG). The TOPAS-C was pumped by 1.8 W of the 800 nm fundamental (4.2 W, 30 fs, 1 kHz) from a Ti-Sapphire regenerative amplifier (Spitfire HP, Spectra Physics). The maximal pulse energy was 20 – 30 μJ, pulse duration was about 100 fs (FWHM) at 1 kHz repetition rate, spectral width (FWHM) — 200 – 250 nm. The diameter of the output beam was about 3 mm ($e^{-2}$ level). Femtosecond laser pulses were focused by a thin $CaF_2$ lens with the 15 cm focal length on the front surface of $CaF_2$ and $BaF_2$ samples of 40 mm length. The beam waist diameter was about 320 μm. The length and location of glowing plasma channels inside the samples were registered by a digital camera Nikon D800 with 0.5 ms exposure time through the polished side face of the samples. With another camera Canon EOS 450D we recorded CE images at a screen located at 10 cm from the sample output surface. Spectroscopic measurements of SC were carried out using the fiber spectrometer ASP-100MF and the original spectrometer ASP-IRHS (Avesta Ltd.) in the spectral ranges 200 – 1100 nm and 1200 – 2600 nm, respectively. For the range 2500 – 4500 nm we used a pyroelectric detector combined with Solar TII monochromator MS2004. In all measurements the entire CE was angularly integrated.

For numerical simulations we used the nonlinear equation for pulse slowly varying envelope. Our model [26,27] includes diffraction, Kerr nonlinearity with stimulated Raman scattering (SRS), plasma nonlinearity, multiphoton absorption and linear attenuation, self-phase modulation (SPM), self-steepening, and full dispersion according to Sellmeier formula. In considered fluorides the GVD of Mid-IR pulses is anomalous as the second-order dispersion coefficient $k_2$ is negative. At the wavelength $\lambda_0 = 3000$ nm for $BaF_2$ $k_2 = -0.04$ fs$^2$/μm and for $CaF_2$ $k_2 = -0.1$ fs$^2$/μm. In the simulations the nonlinear coefficients $n_2 = 2.85 \times 10^{-20}$ m$^2$/W and $n_2 = 1.9 \times 10^{-20}$ m$^2$/W were taken for $BaF_2$ and $CaF_2$, respectively [28]. Corresponding values of the critical power for self-focusing $P_{cr}$ are equal 32 MW and 50 MW. The $BaF_2$ and $CaF_2$ band gaps are close — 9.1 eV and 10 eV and orders of multiphoton ionization for $\lambda_0 = 3000$ nm are equal to $K = 22$ and $K = 25$, respectively. The main pulses parameters were close to those used in the experiment. The peak power of laser pulses was slightly more than $P_{cr}$ and was varied in the range (1.4 – 3.5) $P_{cr}$.

## 3. Conical emission and plasma channels

After exceeding the critical power the appearance of glowing filament areas within the samples and, simultaneously, colored CE rings in the far field at the sample exit was registered. In $BaF_2$ the filament formation starts when the 3000-nm pulse energy is greater than 4 μJ. For 8 μJ pulses ($P \approx 2.5 P_{cr}$) two separated areas of luminescence of 1.0 – 1.5 mm length form in the sample [Fig. 1(a)]. It could be seen that this luminescence is blue in contrast to red luminescence observed for $CaF_2$ in Fig. 2(a). The origin of blue color is $BaF_2$ fluorescence peaked at 330 nm and extended towards 450 nm [5], which is more bright than red plasma recombination radiation currently observed, for instance, in fused silica [19]. In $CaF_2$ the filamentation of 3000-nm pulse arises at the energy of 8 μJ ($P \approx 1.5 P_{cr}$). When the pulse energy reaches 16 μJ ($P \approx 3.1 P_{cr}$) a chain of glowing areas with the overall length up to 15 mm forms in the sample [Fig. 2(a)]. For $CaF_2$ fluorescence spectrum is peaked at 210 nm, so it is invisible in comparison to red recombination radiation. With pulse energy increasing the distance to the filament start and the interval between areas of luminescence decreases. This effect is characteristic for both normal and anomalous GVD regime [16,17,29].

The angular divergence of the CE spectral components $I(\theta, \lambda)$ increases with decreasing of their wavelength $\lambda$, as in the case of near-IR pulse filamentation in normal GVD regime [3]. Divergent components of CE from $BaF_2$ have a rich yellow-green color [Fig. 1(c)]. The $CaF_2$ CE represents green-blue concentric rings [Fig. 2(c)] indicating that the blue shift of the SC visible wing in $CaF_2$ filament is greater than in $BaF_2$. It can be seen that the numerical images of CE rings, processed in accordance to the spectral sensitivity of the camera Canon EOS 450D in the CIE RGB color space [Fig. 1(d) and Fig. 2(d)], coincide very well with the experimental patterns.

The spatial distributions of the electron density $N(r, z)$ in the filament obtained from numerical simulations [Fig. 1(b) and Fig. 2(b)] closely reproduce the picture of the glowing areas registered in the experiment [Fig. 1(a) and Fig. 2(a)].

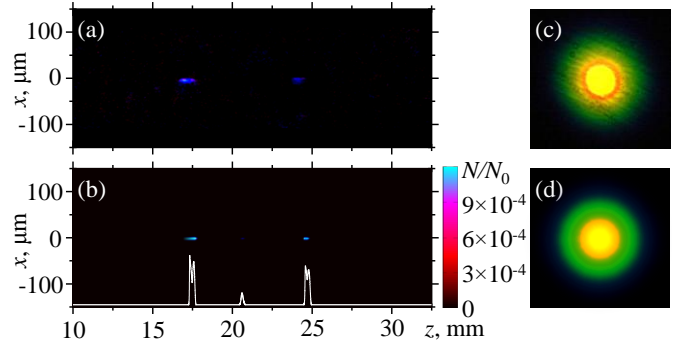

Fig. 1. Experimental images of glowing plasma channels within $BaF_2$ sample (a) and numerical distributions of the electron density $N(r, z)$ (b). The white profile — the electron density distribution on the beam axis. Experimental (c) and numerical (d) patterns of CE visible rings. 3000-nm pulse energy is 8 µJ ($P \approx 2.5 P_{cr}$).

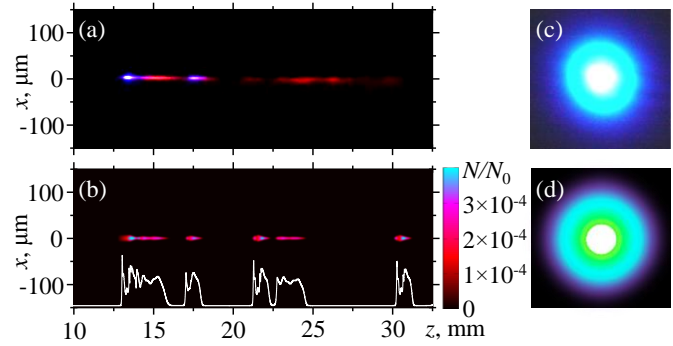

Fig. 2. Experimental images of glowing plasma channels within $CaF_2$ sample (a) and numerical distributions of the electron density $N(r, z)$ (b). The white profile — the electron density distribution on the beam axis. Experimental (c) and numerical (d) patterns of CE visible rings. 3000-nm pulse energy is 16 µJ ($P \approx 3.1 P_{cr}$)

## 4. 3000-nm light bullet

Generation of plasma channels and broadening of the frequency-angular spectrum are directly related to the transformation of the spatiotemporal intensity distribution in the filament. In Fig. 3 numerically calculated evolution of the light field intensity $I(r, t)$ of 3000-nm pulse with 16 µJ energy at several characteristic distances in $CaF_2$ is shown in logarithmic scale colors. The filament is started at $z_f = 13$ mm, where the pulse peak intensity reaches the value $I = 5 \times 10^{13}$ W/cm$^2$ that is enough for multiphoton and avalanche ionization of the medium and results in laser plasma formation [Fig. 3(a) and Fig. 2(b)]. In the strong anomalous GVD regime the light field intensity increases due to both spatial self-focusing and temporal pulse self-compression, which leads to the formation of a high intensity LB [15–17]. Recombination luminescence of laser plasma created by LB was registered in the experiment. The aberration defocusing in laser plasma leads to formation of a ring structure in the LB tail [Fig. 3(b)]. The strong anomalous GVD compresses the LB and its duration becomes less than 10 % of the pulse initial duration. At the same time, a pulse tail refocusing starts and its temporal slices, which have been defocused in plasma, tighten again to the beam axis [Fig. 3(c)]. In contrast to the filamentation in a normal GVD regime, in conditions of anomalous GVD the light field refocusing in space is accompanied by a temporal compression. As a result, at the distance $z = 17$ mm a new LB appears [Fig. 3(d)]. This process can be repeated many times. Each increasing of the pulse intensity is accompanied by a surge of free electrons density in the distribution $N(r, z)$ that was registered as a chain of glowing areas in the filament [Fig. 2(a)].

By measuring the glowing area length, we estimated the lifetime of a LB in the sample. For the $1-5$ mm plasma channel the 3000 nm-pulse LB lifetime is about $5-25$ ps.

The sharpening of the LB trailing edge slope due to self-steepening and defocusing in laser plasma leads to a strong phase modulation of the light field in time and, as a consequence, to a strong anti-Stokes broadening of the pulse spectrum. On increasing the multiphoton order $K$ a steepness of the LB trailing edge increases that causes an enrichment of blue-shifted spectral band [11,12,30].

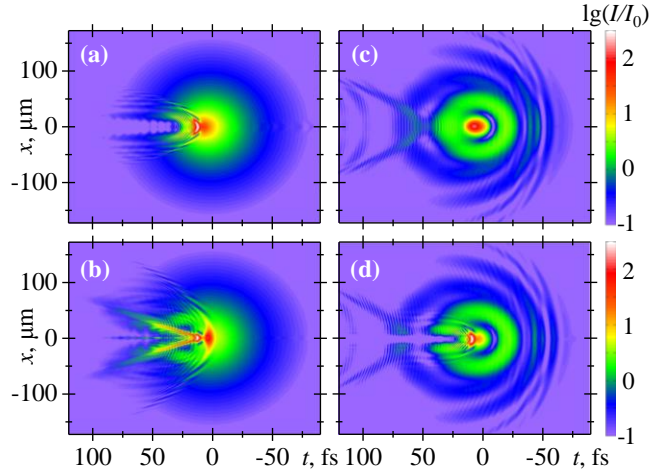

Fig. 3. Spatiotemporal intensity distribution $I(r, t)$ of 3000 nm-pulse in $CaF_2$ at several distances $z$: (a) $z = 13.1$ mm, (b) $z = 13.7$ mm, (c) $z = 16.7$ mm, (d) $z = 17.0$ mm. Pulse energy — 16 µJ, initial pulse duration — 100 fs (FWHM), $I_0 = 3.6 \times 10^{11}$ W/cm$^2$

## 5. Visible band

The experimental and numerical spectra $I(\lambda)$ of 3000-nm pulse after filamentation in $BaF_2$ and $CaF_2$ within the spectral range of $300-4500$ nm are shown in Fig. 4(a) and 4(b). Numerical spectra $I_{num}(\lambda)$ are obtained from computer simulations of frequency-angular spectra $I_{num}(\theta, \lambda)$: $I_{num}(\lambda) = \int I_{num}(\theta, \lambda) d\theta$. Experimental angularly integrated spectra $I_{exp}(\lambda)$, measured in the infrared band, are normalized to the maximum of the corresponding numerical spectral components $I_{num}(\lambda)$ in the range $1300-4500$ nm, the spectra measured in the visible band are normalized to the maximum of $I_{num}(\lambda)$ in the range $300-1100$ nm.

The new effect in Mid-IR filamentation is the extralarge blue shift of an intense SC isolated band in the visible spectral range. For the filamentation of 3000-nm pulse the maximum of the narrow band is located at 570 nm in $BaF_2$ [Fig. 4(c)] and at 520 nm in $CaF_2$ [Fig. 4(d)]. The spectral width (FWHM) of this band is 70 nm for $BaF_2$ and 50 nm for $CaF_2$.

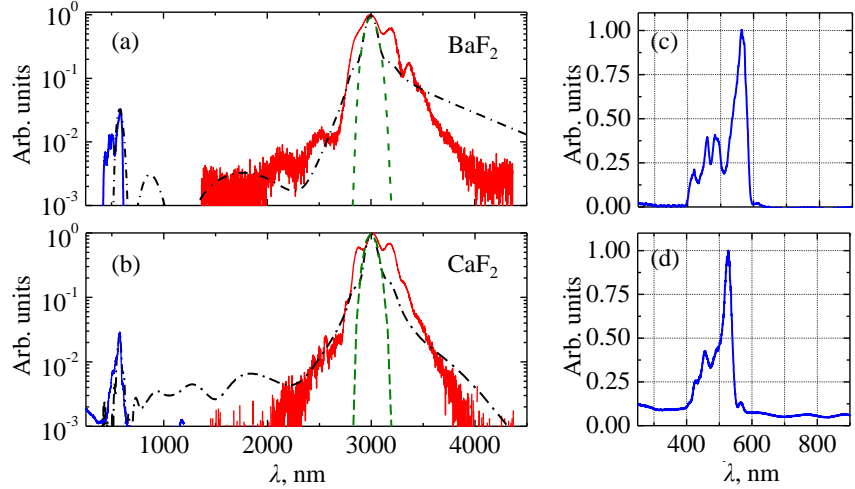

Fig. 4. 3000-nm pulse spectra after filamentation in $BaF_2$ (a,c) and $CaF_2$ (b,d). Solid line — experimental results $I_{exp}(\lambda)$, dash dot line — numerical simulation $I_{num}(\lambda)$. Dashed line — the pulse initial spectrum

SC short-wavelength cutoff in $CaF_2$ is less than that in $BaF_2$ due to higher multiphoton order $K$. Such a dependence of cutoff on $K$ has been revealed and investigated in [12] under filamentation of near-IR pulse in fused silica. The value of the visible band maximum reaches $3 \times 10^{-2} I(\lambda_0)$, whereas the level of the wide separating plateau between the central wavelength $\lambda_0$ and the visible band does not exceed $5 \times 10^{-3} I(\lambda_0)$, where $\lambda_0 = 3000$ nm. The total conversion efficiency from Mid-IR to the visible band exceeds 1 % in both media. The spectral broadening around the fundamental wavelength is also observed [Fig. 4] in contrast to [24].

It should be noted that a characteristic peak centered at 660 nm with intensity of less than $10^{-4} I(\lambda_0)$ has been observed under filamentation of 2000 nm femtosecond pulse in $CaF_2$ [10] and ascribed to third harmonic (TH) generation. Increasing of 2000 nm pulse intensity resulted in TH spectrum monotonous broadening to the anti-Stokes region contrary to our results for 3000 nm pulse. Our experimental spectra are in good agreement with numerical ones calculated under account of phase self-modulation due to Kerr nonlinearity with stimulated Raman scattering (SRS), plasma nonlinearity, multiphoton absorption, linear attenuation, self-steepening, and full dispersion [Fig. 4]. Some discrepancy between the predicted and experimental data in the region 1000 – 2500 nm is due to rather low signal to noise ratio for measurements in this spectral range and the deficiency of a spectrometer which can work well in the spectral range 1000-1200nm.

Increasing the initial pulse central wavelength $\lambda_0$ we observed an increase of the anti-Stokes shift and spectral narrowing of the isolated visible band [Fig. 5]. At the wavelength $\lambda_0 = 3800$ nm the spectral maximum of the visible band blueshifts up to $\lambda_C = 400$ nm in $BaF_2$ and to 330 nm in $CaF_2$. The bandwidth in the last case reduces to 14 nm (FWHM) that is less than the bandwidth of the initial laser pulse. A similar effect was registered for filamentation of near-IR pulses at central wavelength tunable in the range 1300 – 2400 nm in fused silica [12–14] and 800 – 1350 nm in water [31]. However, for investigating in the present experiments Mid-IR pulse filamentation the effect of spectral narrowing of the isolated visible band is more pronounced. The main reason of the observed effect is an increase of multiphoton order $K$ and parameter $|k_2|$, which determines anomalous GVD influence on the process, as in the above mentioned case of the near IR filamentation.

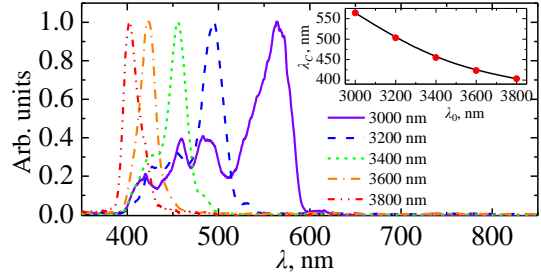

Fig. 5. Experimental spectra of an isolated visible wing of femtosecond pulse filament in BaF$_2$ at central wavelength $\lambda_0$ tunable from 3000 to 3800 nm (left). The top inset — visible wing maximum location $\lambda_C$ depending on the pulse central wavelength $\lambda_0$ (right).

## 6. Interference model

For analytical study of the isolated visible band formation we used the interference model [32] that allows to determine a frequency-angular distribution of the SC spectral components. This model is based on coherence of SC [33], which forms due to the LB field self-modulation. According to the interference model the CE is a result of the interference of the light field of a moving broadband source that is located on the tail slope of the LB, which moves with pulse group velocity $v_g$. The phase shift for the SC component at a wavelength $\lambda$ emitted by point source at an angle θ is:

$$\Delta\phi(\theta,\lambda) = \frac{2\pi l}{\lambda_0}\left(\left(1-\frac{\lambda_0}{\lambda}\right)\frac{c_0}{v_g} - \left(1-\frac{\lambda_0 n(\lambda)}{\lambda n_0}\cos\theta\right)n_0\right), \quad (1)$$

where $n_0 = n(\lambda_0)$, $l$ is the length where the LB exists. The material dispersion of the medium is represented by the dependence of the refractive index $n(\lambda)$ (for example from [28]). The frequency-angular pulse spectrum can be obtained from:

$$I(\theta,\lambda) = I_0(\theta,\lambda) l^2 \text{sinc}^2\left(\frac{\Delta\phi(\theta,\lambda)}{2}\right), \quad (2)$$

where $I_0(\theta, \lambda)$ is a SC spectral amplitude of the LB. In this case the zero and next orders maxima in the frequency-angular distribution $I(\theta, \lambda)$ (2) of SC are determined from the condition of the constructive interference of broadband radiation. In alternative three-wave mixing model [34] only the SC zero-order maxima are determined by the phase matching condition of nonlinear interaction of radiation in thin filament. The interference model describes angle splitting of CE rings under light pulse refocusing in filament in accordance with experiment [19,35].

Figure 6 shows that the interference model reproduces the basic features of the spectra obtained by numerical simulation. Isolated band of visible light is given by CE, which divergence increases with a shorter wavelength. The spectrum visible band in CaF$_2$ is more narrow and its blue shift is greater than in BaF$_2$ (see also Fig. 4). In a narrow visible band there is a constructive interference of SC forming a divergent radiation of CE. Along with zero order interference maximum there are weaker next orders interference bands which are reproduced in numerical simulations and analytic model. In the spectral range between the visible wing and the neighborhood of the central wavelength of initial Mid-IR pulse the light field interference as a whole is destructive and the local maxima are not intense. Our calculations on the base of interference model show that the destructive interference domain broadens with wavelength increasing (and, vice-versa, constructive interference domain becomes narrower). These calculations are completely confirmed by experiment The agreement of the SC frequency-angular spectrum calculated from the interference model with numerical and experimental results has been discussed in detail [19] for filamentation of various wavelengths laser pulses in fused silica.

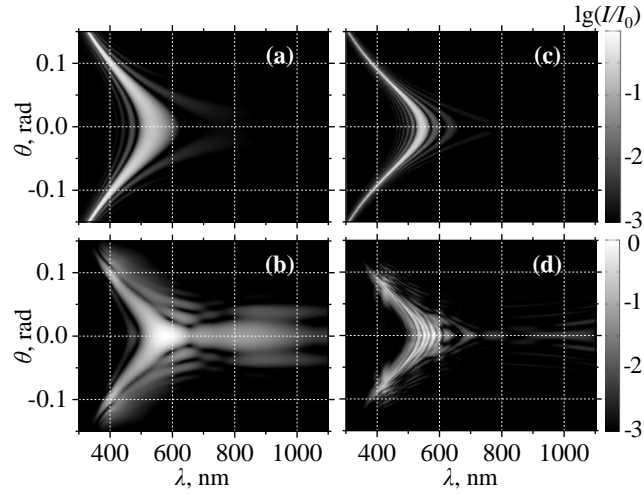

Fig. 6. SC frequency-angular spectra in the visible band, calculated from the interference model (a), (c) and obtained from numerical simulation (b), (d) for 3000-nm pulse filamentation in $BaF_2$ (a), (b) and $CaF_2$ (c), (d)

## 7. Conclusion

Generation of visible light in SC and plasma channels luminescence in $BaF_2$ and $CaF_2$ is investigated in single filament regime of Mid-IR femtosecond laser pulse. For laser wavelength more than 3000 nm, which is in the anomalous GVD region for these materials, plasma channels forms high intensive LBs in the result of spatial and temporal compression of the laser pulse due to phase self-modulation. Several LBs are sequentially arising due to spatial and temporal refocusing of the pulse in a filament. For the first time in Mid-IR spectral region more than 3000 nm blue shift of an isolated visible band of SC in $BaF_2$ and $CaF_2$ femtosecond filament was registered and its change on the variation of the laser pulse wavelength was investigated. For 3000-nm pulse SC visible band is located at 570 nm for $BaF_2$ and 520 nm for $CaF_2$, its bandwidth are 70 and 50 nm (FWHM) correspondingly. Increasing the initial pulse central wavelength to 3800 nm results in blue shift increasing and visible bandwidth reduction so that visible SC wavelengths become 400 nm and 330 nm for $BaF_2$ and $CaF_2$, respectively, its spectral width being reduced to 14 nm (FWHM). Larger shortwavelength shift and bandwidth reduction for $CaF_2$ in comparison to $BaF_2$ is due to larger values of $k_2$ (2.5 more times in the first) and multiphoton order $K$. It was discovered that the formation of this narrow band of visible light is explained as a result of constructive interference of broadband SC emitted by a moving LB in filament under anomalous GVD. The low intense broad spectral band in the spectral range between the visible wing and the central wavelength of initial Mid-IR pulse is formed as a result of destructive interference of SC.

## Acknowledgments

This work is supported by grant RFBR 14-22-02025-ofi_m and The Grant of the President of the Russian Federation NSh-3796.2014.2.